\documentclass[12pt,preprint]{aastex}
\usepackage[twocolumn]{emulateapj5}
\usepackage{apjfonts}
\usepackage{graphicx}

\slugcomment{To be published in the Astrophysical Journal Letters}
\shorttitle{Evidence of the common envelope phase of V471 Tau}
\shortauthors{J.J.~Drake \& M.J.~Sarna}
\begin{document}
\title{X-ray Evidence of the Common Envelope Phase of V471 Tauri}
\author{Jeremy J. Drake\altaffilmark{1} and Marek J. Sarna\altaffilmark{2}}
\affil{$^1$Smithsonian Astrophysical Observatory,
MS-3, \\ 60 Garden Street, \\ Cambridge, MA 02138}
\email{jdrake@cfa.harvard.edu}
\affil{$^2$N. Copernicus Astronomical Center, Polish Academy of Sciences,\\
ul. Bartycka 18, 00-716 Warsaw, Poland}
\email{sarna@camk.edu.pl}

\begin{abstract}

{\it Chandra} Low Energy Transmission Grating Spectrograph
observations of the pre-cataclysmic binary V471~Tau have been used to
estimate the C/N abundance ratio of the K dwarf component for the
first time.  While the white dwarf component dominates the spectrum
longward of 50~\AA, at shorter wavelengths the observed
X-ray emission 
is entirely due to coronal emission from the K dwarf.  The H-like
2p~$^2$P$_{3/2,1/2}$ $\rightarrow$ 1s~$^2$S$_{1/2}$ resonance lines of
C and N yield an estimate of their logarithmic abundance ratio
relative to the Sun of [C/N]$=-0.38\pm 0.15$---half of the
currently accepted solar value.  We interpret this result as the first
clear observational evidence for the presumed common envelope phase of
this system, during which the surface of the K dwarf was contaminated
by CN-cycle processed material dredged up into the red giant envelope.
We use the measured C/N ratio to deduce that 0.015-$0.04\;M_\odot$ was
accreted by the K dwarf while engulfed, and show that this is
consistent with a recent tentative detection of $^{13}$C in the K
dwarf photosphere, and with the measured Li abundance in the scenario
where the red giant companion was Li-rich during the common envelope
phase.

\end{abstract}


\section{Introduction}
\label{s:intro}

The eclipsing Hyades binary V471~Tauri lies at a distance of $47\pm 3$~pc
(Perryman et al.\ 1997) and comprises a cool main-sequence (MS)
star (dK2) and a hot degenerate white dwarf (DA1.5) in an orbit with a
period of 0.52 days.  It is a detached binary in which neither star
fills its Roche lobe, and has a separation of $3.1 R_\odot$
(Young \& Nelson 1972).  The system is evolving toward a cataclysmic
variable phase in which the K dwarf will overfill its Roche lobe and
transfer mass through accretion onto the white dwarf.  As such,
V471~Tau is a rare example of a nearby pre-cataclysmic system and
presents a valuable opportunity to study the origin and evolution of this
type of close binary.

Based on evolutionary time--scales, Vauclair (1972) argued that
the current mass of the V471~Tau system is much lower than that
expected from a progenitor system in which one component has
reached the white dwarf phase.  Moreover, the orbital separation
is much smaller than the radius of the primary star 
during its red giant phase.  It was therefore proposed that
the system had gone through a common envelope (CE) phase, during
which the initial orbital separation was sufficiently large that
the secondary was only engulfed after the primary had attained red
giant dimensions (Paczy\'nski 1976).  During a CE phase, friction
leads to a rapid spiraling down of the orbit.  If the CE is spun
up sufficiently during this time, a substantial fraction can be
ejected from the system thereby avoiding coalescence of the
companion with the red giant core (Iben \& Livio 1993).

Despite strong theoretical arguments in favour of this
evolutionary scenario for V471~Tau and other systems in similar
evolutionary stages, unequivocal observational evidence supporting
the CE phase has so far been lacking.  During a CE phase, some
fraction of the red giant envelope should have been accreted by
the K dwarf.  This phase would therefore be betrayed by the
presence of the products of interior nuclear burning dredged-up
into the envelope of the red giant during the ascent of the giant
branch.

During the MS evolution of a solar-type star, the central
abundance of C decreases to about 5\%\ of its initial value while the
N abundance increases such that the total number of C and N nuclei remains
approximately constant.  The isotopic ratios $^{12}$C/$^{13}$C and
$^{16}$O/$^{17}$O also decrease by about an order of magnitude.
Determining the C and N abundances in the red dwarf from atomic lines
in photospheric spectra is complicated by rotational smearing and the
weakness of available lines of N~I, and, to
our knowledge, no such measurements yet exist for the dwarfs in
pre-cataclysmic systems.  Sarna et al.\ (1995) proposed using the CO
bands near 1.59, 2.3 and 4.6~$\mu$m for determining the
$^{12}$C/$^{13}$C and $^{16}$O/$^{17}$O abundance ratios.
Dhillon et al.\ (2002; see also Catal\'an et al.\ 2001) 
analysed CO band spectra of V471~Tau and
the cataclysmic variable SS~Cyg and found some evidence for the
$^{13}$C expected from the CN-cycle, though the results were
inconclusive because of possible blending with neutral species.

In this paper, using {\it Chandra} Low Energy Transmission Grating
spectrograph (LETGS) X-ray spectra of Ly$\alpha$ resonance transitions
in C and N H-like ions, we determine for the first
time the C/N abundance ratio in the K dwarf secondary of V471~Tau.  
We show that the secondary star has half the
C/N ratio expected for the surface of an unevolved star, presenting
the first definitive evidence of the CE phase in which substantial
surface contamination by the red giant envelope of its companion
occurred.  

\section{Observations and Analysis}
\label{s:obs}

V471~Tau was observed for approximately one day by {\it Chandra} using
the Low Energy Transmission Grating and High Resolution Camera (LETGS)
in standard instrument configurations, between 2002 January~24
UT~22:31 and January~25 UT~23:38.  Observational data were obtained
from the public Chandra Data
Archive\footnote{http://asc.harvard.edu/cda}, and pipeline-processed
(CXC software version 6.3.1) photon event lists were reduced using the
CIAO software package version 2.2.  This latter processing included
filtering of events based on observed signal pulse-heights to reduce
background (Wargelin et al., in preparation).  The resulting spectra
had an effective exposure time of 87494s.  

Spectra were analysed using
the PINTofALE IDL\footnote{Interactive Data Language, Research Systems
Inc.} software suite (Kashyap \& Drake 2000).  The line strengths
analysed here were measured by
fitting ``modified Lorentzian'' functions of the form
$F(\lambda)=a/(1+\frac{\lambda-\lambda_0}{\Gamma})^\beta$, where $a$
is the amplitude and $\Gamma$ a characteristic line width.  For a
value of $\beta=2.4$, it has been found that this function represents
the line response function of the LETG+HRC-S instrument to the
photometric accuracy of lines with of order a few thousand counts or less
(Drake et al., in preparation).

The LETGS spectrum of V471~Tau is characterized at long wavelengths
($> 50$~\AA) by the strong thermal continuum from the white dwarf
photosphere that is expected on the basis of earlier EUV and X-ray
observations (e.g.\ Barstow et al.\ 1992).  However, this continuum
contributes no flux shortward of 50~\AA, and at shorter wavelengths
the spectrum is dominated by easily discernible emission lines that we
interpret as coronal emission from the K dwarf.  No modulation of the
coronal emission with orbital phase was seen, such as might be
expected if the UV radiation field of the white dwarf were 
influencing the corona facing it through modification of its 
underlying chromospheric structure (D.~Garcia et al., in preparation). 

\begin{figure}
\epsscale{1.0}
\plotone{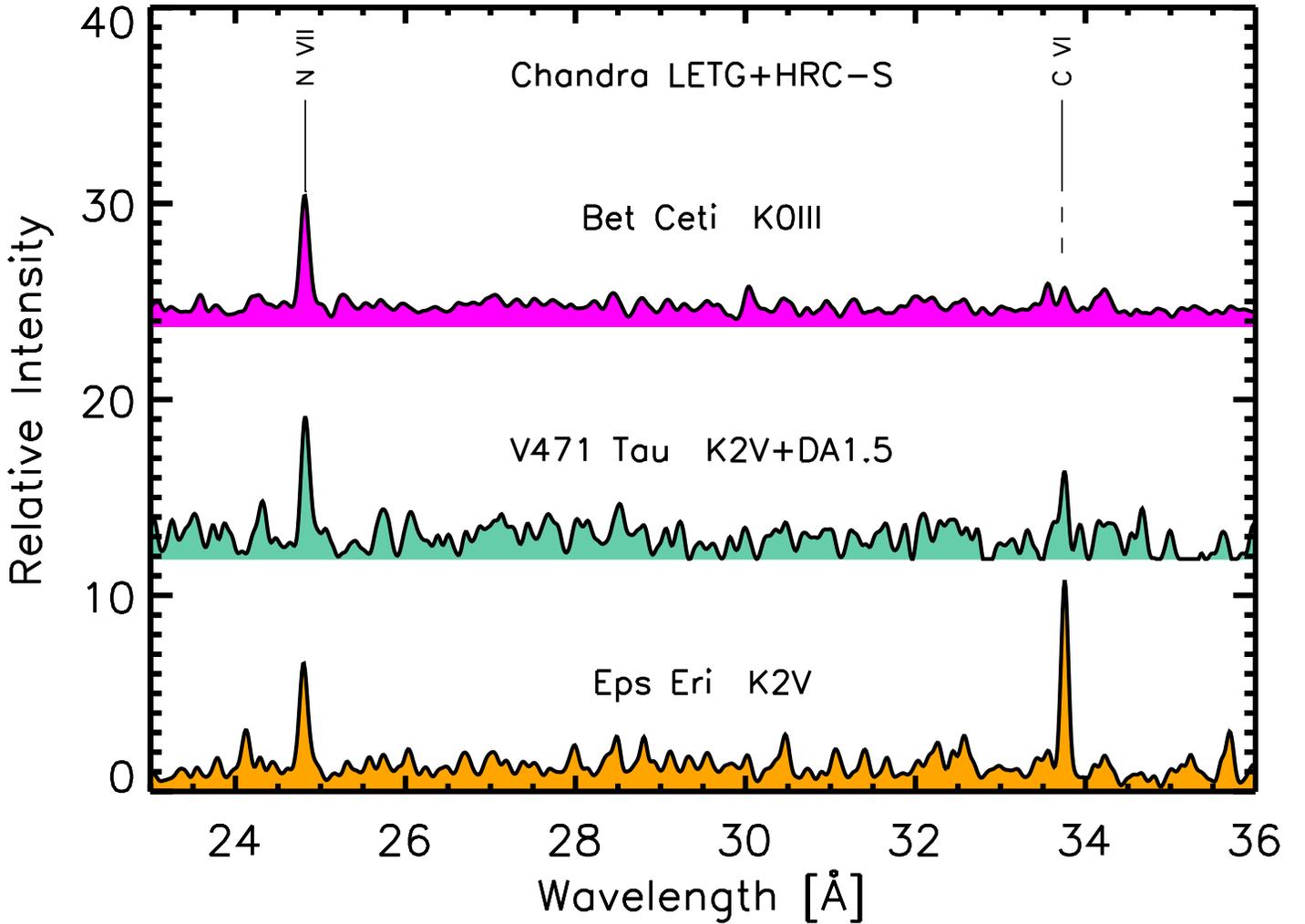}
\caption{
{\it Chandra} LETGS spectra of the evolved K0 giant
$\beta$~Ceti, the
K2 dwarf of the pre-cataclysmic binary V471~Tau, 
and the K2 dwarf $\epsilon$~Eri, covering the range 23-36~\AA .  The
prominent lines are 
the  2p~$^2$P$_{3/2,1/2}$ $\rightarrow$ 1s~$^2$S$_{1/2}$ resonance
transitions of H-like C and N, and the spectra have been normalised
such that the strength of the N line is the same in each.
The ratio of the intensities of these
lines, $I_C/I_N$, is independent of temperature and is directly
proportional to
the C/N abundance ratio.  The C/N abundance ratio in
V471~Tau is intermediate between that of the unevolved dwarf and
evolved giant.}
\label{f:spectra}
\end{figure}

The spectrum in the region between 23-36~\AA\ that
contains the Ly$\alpha$ resonance lines of H-like N and C is
illustrated in Figure~\ref{f:spectra}.  We show in the same figure the
{\it Chandra} LETGS spectra of the K2 dwarf $\epsilon$~Eri and the
K0III giant $\beta$~Ceti obtained from the public archive and
processed in the same way as the V471~Tau observation.  We draw
particular attention to the different relative strengths of the C and
N lines in these stars.  It has recently been shown that the ratio of
the strengths of these lines is essentially independent of temperature
over the range $6.35 \leq \log T \leq 7.4$
(Schmitt \& Ness 2002, Drake 2003a, Drake et al., in preparation). 
 The ratio of their
measured fluxes for active stars with dominant coronal temperatures
$\ga 2\times 10^6$~K, such as the stars in Figure~\ref{f:spectra},
therefore yields directly the C/N abundance ratio.  That all three
stars are indeed dominated by temperatures above $2\times 10^6$~K is
underscored by the relative strengths of the H-like and He-like
resonance lines of N at 24.78~\AA\ and 28.79~\AA, the latter of which
is almost absent.
Since all three
spectra in Figure~\ref{f:spectra} were obtained with the same
instrument, the variation in the relative C and N line strengths,
$I_C/I_N$, can
be interpreted directly in terms of the ratio of the C and N number 
densities, i.e.
\begin{equation}
  \frac{N(C)}{N(N)}= \frac{I_C \epsilon_N(T)}{I_N \epsilon_C(T)}; 
\;\;\;\; 6.35 \leq \log T \leq 7.4 
\label{e:abunrat}
\end{equation}
where $\epsilon_X(T)$ is the photon emissivity of element X under
coronal equilibrium conditions.  In photon units, the PINTofALE 
emissivity ratio in equation~\ref{e:abunrat}, computed using
the CHIANTI~v3 (Dere et al.\ 2001) implementation of the collision
strengths of Aggarwal \& Kingston (1991), together with the ionization
balance of Mazzotta et al.\ (1998), is $\epsilon_N/\epsilon_C=1.84$
(see also Drake 2003a).  

The C and N lines of the unevolved dwarf $\epsilon$~Eri and the
giant $\beta$~Ceti were analysed recently by Schmitt \& Ness (2002).
These authors concluded that the C/N line strength ratio for
$\epsilon$~Eri was consistent with a solar, or ``unevolved'', C/N
abundance ratio.  For the solar photospheric C abundance
C/H=8.39 (Allende~Prieto et 
al.\ 2002) and N abundance N/H=7.92 (Grevesse \& Sauval 1998), the
line intensity ratio is $I_C/I_N=1.6$.
Schmitt \& Ness (2002) measured a value for
$\epsilon$~Eri of $2.1\pm 0.24$.  There appear to be no recent
determinations of the photospheric N abundance in $\epsilon$~Eri,
though the C abundance is within 0.1~dex or so of the solar value
(Takeda et al.\ 2001; Zhao et al.\ 2002).  Allowing for an uncertainty
of 0.1~dex in the photospheric ratio, this is then consistent with the
coronal C/N line ratio.

In contrast, the Schmitt \& Ness (2002) ratio for $\beta$~Ceti,
whose surface composition has been modified by the dredged-up products
of CN-cycling, is $I_C/I_N=0.16\pm 0.05$---an order of magnitude
smaller.  Based on the Drake (2003a) theoretical C/N line strength
ratio for a solar C/N abundance ratio, the measured ratio for
$\beta$~Ceti corresponds 
to an abundance ratio of [C/N]$=-1$ relative to the
Sun.\footnote{Except where stated, we adopt the common notation to
express abundances in which X/H=$\log_{10}[N(X)/N(H)]$+12, where
$N(X)$ is the number density of element X, and [X/Y] represents the
logarithic X/Y abundance ratio relative to the solar value.}  We note
that this is in good agreement with the photospheric measurement of
Lambert \& Ries (1981).

The V471~Tau C and N lines in
Figure~\ref{f:spectra} are in a proportion intermediate between that
of an unevolved composition exemplified by $\epsilon$~Eri and the
post-dredge-up surface composition of $\beta$~Ceti.
A formal value for the C/N abundance ratio of V471~Tau is given by the
observed spectral line flux ratio.  Measured line strengths are listed in
Table~\ref{t:fluxes}.  The intensity ratio to abundance ratio
conversion factor 
from equation~\ref{e:abunrat}, $N(C)/N(N)= 1.85 I_C/I_N$,
leads directly to the result C/N$=0.1\pm 0.15$ (corresponding to
$N(C)/N(N)=1.22\pm 0.36$), or [C/N]$=-0.38\pm 0.15$.

Does this coronal value represent the composition of underlying
atmospheric layers?  A large body of observational evidence now exists
demonstrating that the abundances of elements in the coronae of stars
can differ substantially from their photospheric values (e.g.\ Drake
2003b).  The differences appear to be driven predominantly by the
element first ionization potential (FIP).  In active stars, coronal
abundances of very high FIP elements such as Ne and Ar appear enhanced
relative to abundances of low FIP elements such as Fe, Mg and Si.
Elements such as C, N and O appear to remain close to photosheric
values.  Drake (2003a) concluded that C and N are not fractionated to
any significant extent in the corona of the active binary V711~Tau, as
expected based on their very similar FIPs.  While we cannot rule out
relative fraction among C and N with certainty, this does not appear
likely.  We therefore consider our derived C/N abundance ratio to be
directly applicable to the deeper regions of the V471~Tau K dwarf.

\section{Discussion}
\label{s:discuss}

\subsection{Accretion during the CE phase}
\label{s:accrete}

The duration of a CE phase is thought to range typically from 
$10^2$ to $10^3$yr, with a maximum of about $10^4$yr (Taam 1989; Iben \&
Livio 1993).  Calculations by Hjellming \& Taam (1991) indicate that
the red dwarf can accrete some of the red giant envelope
(0.01-0.2$~M_\odot$) during this time.  Our C/N abundance ratio
enables us to estimate this amount of accreted mass.


For the case of a typical intermediate mass star (2.5$~M_\odot$,
$Z=0.03$), stellar evolutionary calculations predict a decrease in the
surface C/N abundance ratio between the zero age main-sequence (ZAMS)
and immediately prior to the first thermal pulse by a factor of
3.75 (e.g.\ El Eid 1994, Girardi et al. 1996).  From Grevesse \&
Sauval (1998) and Allende Prieto et al. (2001, 2002), the C/N ratio
for the Sun expressed in terms of mass fraction is 2.54 and the mass
fractions of C and N are $X^C_\odot=0.00295$ and $X^N_\odot=0.00116$,
respectively.  Assuming both progenitors of the V471~Tau system had an
initial C/N ratio similar to that of the Sun, our abundance ratio
derived from the {\it Chandra} LETGS spectrum implies a reduction in
the red dwarf C/N ratio by a factor of 2.4

\begin{figure}
\epsscale{1.0}
\plotone{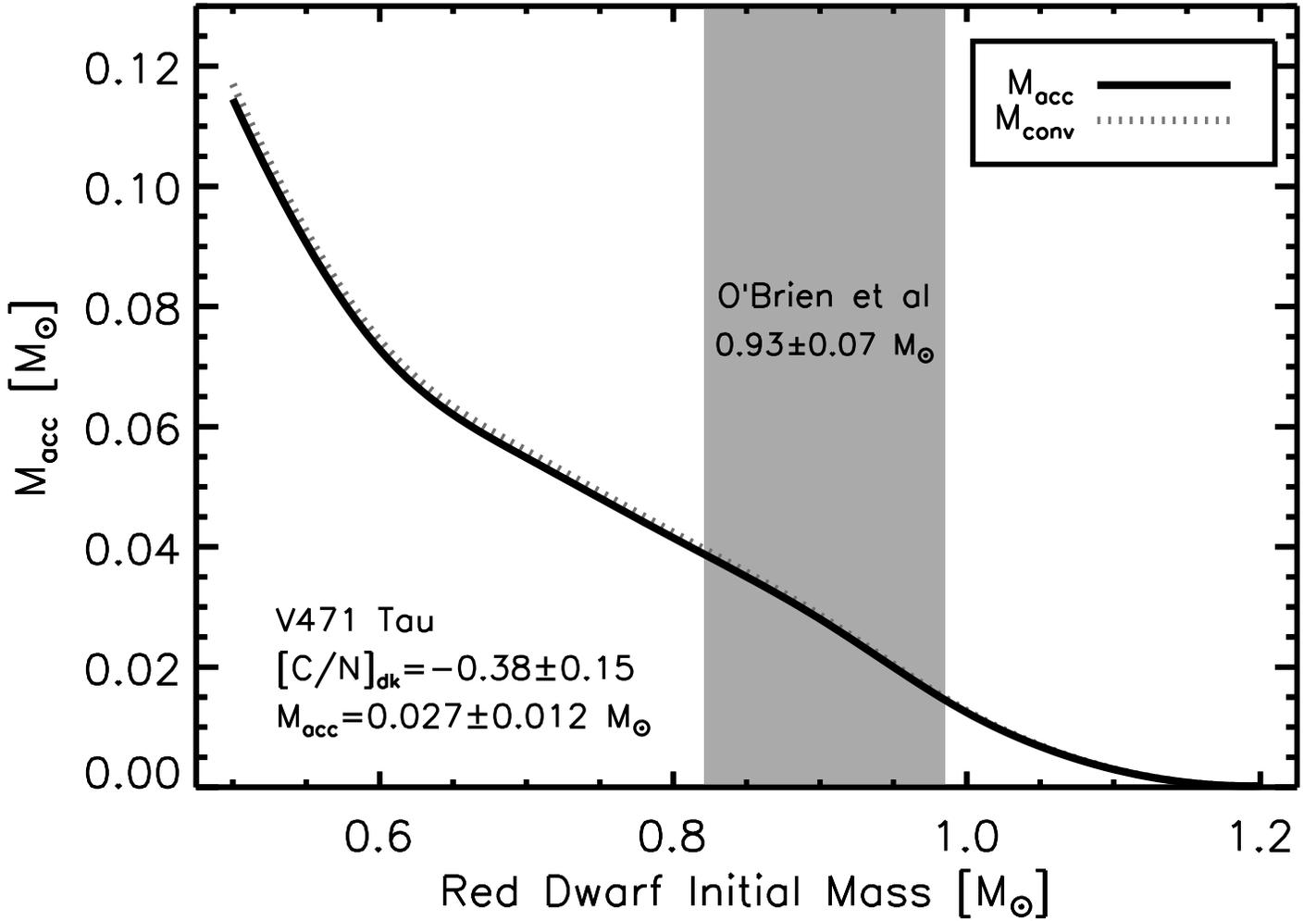}
\caption{The mass accreted onto the K dwarf during the CE phase as a
function of its initial mass, based on the derived C/N abundance
ratio.  The present day mass estimate of
O'Brien et al.\ (2001), corrected for the accreted component, is
illustrated by the shaded region.  We also illustrate the locus
corresponding to the convection zone mass.}
\label{f:macc}
\end{figure}

During the CE phase, the red dwarf secondary effectively exists within
the envelope of the red giant primary.  The accreted material
therefore has the chemical composition of a giant star.  The material
accreted during the CE phase will be mixed on a convective-mixing
(dynamical) time-scale. The mass of the region that is mixed,
$M_{mix}$, consists of the sum of the masses of the convective region
of the original star, $M_{conv}$, and the accreted matter $M_{acc}$.
The C/N abundance by mass fraction at the surface of the red dwarf in
the post-CE phase, $X^C_{rd}/X^N_{rd}$,  is then given by
\begin{equation}
\frac{X^C_{rd}}{X^N_{rd}}
= \frac{M_{conv} X^C_\odot + M_{acc} X^C_{giant}}{
M_{conv} X^N_\odot + M_{acc} X^N_{giant}}.
\label{e:rat}
\end{equation}
The accreted mass can then be written
\begin{equation}
M_{acc} = M_{conv} \frac{X^C_\odot -
(X^C_{rd}/X^N_{rd}) X^N_\odot}{(X^C_{rd}/X^N_{rd})
X^N_{giant} - X^C_{giant}}.
\label{e:macc}
\end{equation}
Here, the C and N mass fractions for the giant phase are assumed
to be the solar values modified such that $X^C_\odot/X^N_\odot$ is
reduced by the factor of 3.75 predicted by evolutionary
calculations, as noted above: $X^C_{giant} $=0.00197,
$X^N_{giant} $=0.0029.  The quantity $M_{acc}$ is illustrated as a
function of the red dwarf initial mass in Figure~\ref{f:macc}. The
convection zone mass, $M_{conv}$, has been taken from models
computed using a standard stellar evolutionary code based on the
Henyey-type method of Paczy\'nski (1970), which has beeen adopted
to low-mass stars (see Marks \& Sarna 1998). The recent mass
determination of O'Brien, Sion \& Bond (2001), $M=0.93\pm
0.07M_\odot$, leads to an accreted
mass in the range 0.015-0.04$~M_\odot $, and a similar initial
convection zone mass.



\subsection{The $^{12}$C/$^{13}$C ratio}

Sarna et al. (1995) described how accretion during the CE
phase changes the $^{12}$C/$^{13}$C ratio on the surface of the red
dwarf.
Analogous to Equation~\ref{e:macc}, the accretion-modified
$^{12}$C/$^{13}$C ratio is given by
\begin{equation}
\frac{X^{12}_{rd}}{X^{13}_{rd}}= \frac{M_{conv} X_\odot^{12} +
M_{acc} X_{giant}^{12}}{M_{conv} X_\odot^{13} + M_{acc} X_{giant}^{13}}
\label{e:c12c13}
\end{equation}
Using the values of $M_{acc}$ and $M_{conv}$ from
\S\ref{s:accrete}, we obtain a mass fraction ratio
$X^{12}_{rd}/X^{13}_{rd} = 30$.  This value is slightly larger than
the tentative estimate of $X^{12}_{rd}/X^{13}_{rd}=10$ based on CO
bands near $2.35\mu$m by Dhillon et al. (2002).

%

%
%
%
%
%
%

\subsection{Lithium abundance}

The Li in a late G or early K dwarf is expected to be highly depleted
by Hyades age, with a value Li/H$\sim 0$ (e.g.\ Barrado~y~Navascu\'es
\& Stauffer 1996; Zboril et al.\ 1997).  However, for Hyades dwarfs in
binaries, Li abundances are higher by up to 3 orders of magnitude; the
largest abundances found are in dwarf members of short period binaries
(Barrado~y~Navascu\'es \& Stauffer 1996).  These high Li abundances
have been interpreted in terms of inhibited mixing in these
rapidly-rotating stars, both during the pre-MS and MS phases (e.g.\
Zahn 1994).  V471~Tau has an abundance Li/H$=2.3$ (Martin, Pavlenko \&
Rebolo 1997; Barrado~y~Navascu\'es \& Stauffer 1996), which is not in
itself conspicuously different to that of similar short period
binaries.  However, the CE scenario and white dwarf age imply that the
close-binary nature of V471~Tau is only very recent.  Prior to this
current evolutionary phase it should have already lost its Li.

Martin et al.\ (1997) suggested that the K dwarf regained Li
from a Li-rich red giant envelope during the CE phase and suggested
that the rare Li-rich giants might be associated with a CE
evolutionary phase.  Li-rich giants have Li/H ranging from $\sim
1.5$-4.75, with a mean from 12 stars of Li/H=2.3 (De Medeiros et al.
1996).  The expression for the present day Li mass fraction in a dwarf
accreting from such a Li-rich giant is
\begin{equation}
X_{rd}^{Li} = {{M_{conv} X_{rd}^{Li} + M_{acc} X_{giant}^{Li}}
\over {M_{conv} + M_{acc}}} \approx {{M_{acc} X_{giant}^{Li}}
\over {M_{conv} + M_{acc}}}
\end{equation}
Based on the observed present day Li abundance and our values of
$M_{acc}$ and $M_{conv}$, we can estimate the required Li abundance of
the red giant envelope to be in the range Li/H$=2.3$-2.8 immediately
prior to the CE phase.  This result is in good agreement with typical
values for Li-rich giants (De Mederios et al.\ 1996, 1997).

Charbonnel \& Balachandran (2000) identify two distinct episodes of Li
production which occur during RGB and early AGB phases, depending upon
the mass of the star. In the case of V471~Tau, the progenitor of the
white dwarf must have been more massive than the red giants now in the
Hyades and presents an evolutionary problem (e.g.\ O'Brien et al.\
2001 and references therein).  Either it was born significantly later
than other Hyades members---a scenario that cannot be ruled out with
certainty---or else it gained a significant amount of mass.  Iben \&
Tutukov (1999) and O'Brien et al.\ (2001) suggest the WD progenitor
might be the evolved remnant of a merged close binary which formed a
blue straggler.  While Ryan et al. (2001) have shown that blue stragglers in
binary systems are lithium-deficient, in this evolutionary scenario
the Li currently observed would have been produced during the
subsequent giant phase.

\section{Conclusions}

We have determined the C/N abundance ratio in the corona of the K
dwarf component of V471~Tau based on the resonance lines of H-like
C and N seen in {\it Chandra} LETGS spectra.  The value we obtain,
C/N$=0.1\pm 0.15$, or [C/N]$=-0.38\pm 0.15$, is intermediate
between that of an unevolved surface composition and the surface
composition of typical red giants.  This result provides the most
direct observational evidence to date of the CE phase of V471~Tau.
Our C/N abundance implies the K~dwarf accreted $\sim 0.015$-$0.04
M_\odot$ from the envelope of the primary star during its red
giant phase. This material was subsequently diluted by mixing with
the convective envelope of the K dwarf.  Our value for the
accreted mass predicts a $^{12}$C/$^{13}$C abundance ratio that is
consistent with a recent observational estimate, and a surface Li
abundance in agreement with observations in the scenario where the
red giant envelope was Li rich during the CE phase.

\acknowledgments

We thank the NASA AISRP for providing financial assistance for the
development of the PINTofALE package, and the CHIANTI project for
making publicly available the results of their substantial effort in
assembling atomic data useful for coronal plasma analysis.  JJD was
supported by NASA contract NAS8-39073 to the {\em Chandra X-ray
Center} during the course of this research.  MJS was supported through grant
5-P03D-004-21 of the Polish State Committee for Scientific Research.




\begin{deluxetable}{llllllll}
\tablecaption{Measured C and N (2p)$^2$P$_{1/2,3/2}$ $\rightarrow$
(1s)$^2$S$_{1/2}$ Fluxes \label{t:fluxes}} 
\tablehead{
\colhead{$\lambda_{\rm obs}$} &
\colhead{Ion} &
\colhead{$\log T_{\rm max}$\tablenotemark{a}} &
\colhead{Intensity} &
\colhead{Area} &
\colhead{Flux/$10^{-14}$}
 \\
\colhead{(\AA )} &
\colhead{} &
\colhead{(K)} &
\colhead{[Counts]}  &
\colhead{[cm$^2$]} &
\colhead{[erg~cm$^{-2}$~s$^{-1}$]}
}
\startdata
$24.800\pm 0.009$ & 
N VII  & 6.30 &
$74\pm 12$ &
15.2 &  $4.46\pm 0.72$\\
$33.725\pm 0.009$ & 
C VI   & 6.20 &
$37\pm 9$ &
11.6 & $2.15\pm 0.52$ \\
\enddata
\tablenotetext{a}{The temperature at which the line emissivity peaks.}
\end{deluxetable}

\end{document}